\documentstyle[prd,aps,epsfig,psfig]{revtex}
\tighten
\begin{document}
\draft
\twocolumn[\hsize\textwidth\columnwidth\hsize\csname
@twocolumnfalse\endcsname
\title{The No-defect Conjecture~: Cosmological Implications}
\author{Jean-Philippe Uzan}
\address{D\'epartement d'Astrophysique Relativiste et de Cosmologie,\\
Observatoire de Paris-Meudon, UPR 176, CNRS, 92195 Meudon (France).}
\date{\today}

\maketitle
\begin{abstract}
When the topology of the universe is non trivial, it has been shown
that there are constraints on the network of domain walls, cosmic
strings and monopoles.  I generalize these results to textures and study
the cosmological implications of such constraints. I conclude that a
large class of multi-connected universes with topological defects
accounting for structure formation are ruled out by observation
of the cosmic microwave background.
\end{abstract}
\vskip2pc]
The question of what generated the initially small cosmological
inhomogeneities responsible for the large scale structures observed
in the universe is still open.  Two main families of models able to
explain the origin of these inhomogeneities are currently
investigated~:the inflationary scenarios
\cite{guth81,steinhardt95} find their origin in  the
amplification, due to accelerated cosmic expansion, of quantum
fluctuations and, in the topological defect (TD) scenarios, they are
seeded by TD which can appear during a spontaneous symmetry breaking
phase transition \cite {kibble76,shellard94}. Inflationary models seem
more promising that currently investigated defect models and future
observations of the anisotropies of the cosmic microwave background
(CMB) radiation by the MAP \cite{map} and Planck \cite{planck}
satellite missions, should discriminate between these scenarios.

When computing the CMB anisotropies in both families of models, it is
usually implicitly assumed that the spatial sections of the universe
are simply connected. There are however many reasons for considering
multi-connected universes (see e.g.  \cite{luminet95,uzan97b}).  Such
``small universes'' \cite{ellis71} have the same local geometry and
dynamics as simply-connected Friedmann-Lema\^\i tre universes but
different {\it global} structure. There are observational constraints
on the characteristic size of such universes from the non detection of
multiple images in cluster catalogs
\cite{lehoucq96}.  In locally euclidean or spherical universes with a 
non trivial topology, there must exist a long wavelength cut-off in
the temperature anisotropies of the CMB. The non-detection of such a
cut-off allows the exclusion of flat universes with multi-connected
spatial sections \cite{stevens93}. This was generalized to a
non-compact, infinite volume hyperbolic topology describing a toroidal
horn \cite{levin97}. This method does not necessarily apply to
universes with compact hyperbolic spatial sections \cite{cornish98}
since they can support perturbations with wavelengths larger than the
curvature scale. Recently, it has been shown that the topology can
have a precise signature on the microwave sky \cite{cornish97} : since
the observed last scattering surface is a 2-sphere, it will intersect
itself along circles if the topology is non trivial. The number, size
and orientation of these matched circles is a signature of the
topology. Most of the conclusions drawn from the CMB rely on the fact
that the perturbations were produced during an inflationary phase. It
is thus of interest to wonder if they hold in presence of TD since
they will go on seeding perturbations and for instance generate low
multipoles by a perspective effect. Let me emphasize that TD appear in
a lot of extensions of the standard model of particle physics and that
they can affect the small CMB multipoles and change the conclusions
drawn on the topology from the study of the CMB even if they cannot
explain the whole CMB spectrum.

In this article, I want to show that in a wide class of multi-connected
universes, TD cannot account for structure formation.
I will concentrate on universes with compact hyperbolic spatial
sections. Such universes have a remarkable property that
links topology and geometry~: the {\it rigidity theorem} 
\cite{mostow73} implies \cite{luminet95} that 
geometrical quantities such as the volume, the lengths of its closed
geodesics etc. are topological invariants. Thus once the topology of
the spatial sections and the cosmological parameters have been chosen,
the cosmological model is completely determined.

I first recall the No-defect conjecture \cite{uzan97} and generalize
it to textures. I then discuss the cosmological implication of this
result. For that purpose I show that there exists a cut-off in the
angular power spectrum of the CMB temperature anisotropies. This
cut-off depends only on cosmological parameters and on a topological
invariant of the spatial sections.
\section{Constraints on the TD Network}
The appearance and the nature of TD during a phase transition
depend only on the topology of the vacuum manifold $G/H$ when a
larger group $G$ is broken down to a smaller group $H$
\cite{kibble76}. If $\pi_0(G/H) \not\sim \lbrace0\rbrace$ then domain
walls might form while cosmic strings, monopoles and textures appear
respectively if $\pi_1(G/H)\not\sim \lbrace0\rbrace$,
$\pi_2(G/H)\not\sim \lbrace0\rbrace$ and $\pi_3(G/H)\not\sim
\lbrace0\rbrace$, $\pi_n$ being the $n^{th}$ 
homotopy group. Moreover, depending on whether the broken symmetry is
local (i.e. gauged) or global (i.e. rigid), the defects will be called
``local'' or ``global'' \cite{shellard94}. For local defects, the
energy is strongly confined and there is a considerable gradient of
energy which implies that there are no long range interactions between
them whereas for global defects the energy is spread out over
typically the horizon scale. I refer to the horizon size as the size
of the {\it particle horizon} \cite{rindler56}, i.e.  of causally
connected regions.

The manifold ${\cal M}$ describing the universe is assumed to be
globally hyperbolic so that ${\cal M}=\Sigma\times R$ and that the
spatial sections $\Sigma$ are orientable, locally homogeneous and
compact. They can then be described by their fundamental polyhedron
${\cal P}$, which is convex and has a finite number of faces
identified by pairs, together with the holonomy group $\Gamma$. For
compact euclidean 3-manifolds (e.g. 3-torus) ${\cal P}$ may possess
arbitrary volume, but no more than eight faces \cite{luminet95}. For
compact hyperbolic 3-manifolds, as stressed in the introduction, the
volume of ${\cal P}$ is a topological invariant. I will denote $r_+$
the outradius, i.e.  the radius of the smallest geodesic ball that
contains ${\cal P}$. These quantities can be found by using the
program {\it SnapPea}
\cite{weeks}.

P. Peter and myself have shown in \cite{uzan97} that if we assume that
the field theory is not twisted \cite{isham78} and if the
characteristic lengths of ${\cal P}$ were smaller than the horizon
size today, the constraints imposed by the global topology on the TD
network imply that extended defects (namely strings and walls) do not
exist today.

This result can be generalized to textures. They are field
configurations that contract and unwind producing points of higher
energy at given points of the spacetime\cite{turok90} . The network of
such defects is not constrained by the topology in the same manner
as monopoles, strings and walls. In a simply-connected universe,
this unwinding of the field configuration is a continuous process.
However, when the topology is non trivial, there is a time after which
all scales have entered the horizon, i.e after which the entire
spatial sections are causally connected. Thus, the finiteness of the
spatial sections implies that there is no texture left and that no
inhomogeneities are seeded after a given time. This result is new
and was not contained in our previous analysis \cite{uzan97}.

\section{Implications on  the CMB}
In this section, I first compute the time, $\eta_{disp}$, after which
all TD have disappeared as well as the angle subtended by the
fundamental polyhedron at $\eta_{disp}$. I then estimate the angular
power spectrum of the CMB temperature anisotropies on large scales and
compare it qualitatively with COBE observations.
\subsection{Lifetime of the TD network}
I restrict myself to Friedmann-Lema\^\i tre universes. The spatial sections
$\Sigma$ are thus homogeneous and isotropic and the line element reads
\begin{equation}
ds^2=a^2(\eta)\left\lbrace-d\eta^2+ 
d\chi^2+\sinh{\chi}^2 \left(d\theta^2+\sin{\theta}^2d\phi^2\right)
\right\rbrace,
\end{equation}
where $a$ is the scale factor and $\eta$ the conformal time.
In a matter dominated universe, $a(\eta)$ is given by
(see e.g. \cite{peebles93})
\begin{equation}
\frac{a}{a_0}=\frac{\Omega_0}{2(1-\Omega_0)}\left(\cosh{\eta}-1\right),
\end{equation}
where $\Omega_0$ is the density parameter and all present time
quantities have a subscript $0$.  The comoving Hubble radius is then
given by
\begin{equation}
R_H^{com}\equiv 1/{\cal H}=\left({\cosh{\eta}-1}\right)/{\sinh{\eta}},
\label{a3}
\end{equation} 
${\cal H}$ being defined as ${\cal H}\equiv a'/a$ and a prime being a
derivative with respect to $\eta$. The Friedmann equation takes the form
\begin{equation}
{\cal H}^2(1-\Omega)=1.\label{fried}
\end{equation}
Comoving lengths
being normalized to units where the curvature radius is unity, the
``physical'' length today is then given by
\begin{equation}
L^{phys}_0\equiv a_0L^{com}=H_0^{-1}(1-\Omega_0)^{-1/2}L^{com},\label{lphys}
\end{equation}
where $H_0={\cal H}_0/a_0$ is the Hubble parameter. I introduce the
redshift $z$ by the relation
\begin{equation}
1+z\equiv a_0/a.
\end{equation}
Assuming that the universe is matter dominated at least since the
last scattering surface means that the equality (subscript $eq$)
occurs before the decoupling (subscript $LSS$), i.e. that
$z_{eq}>z_{_{LSS}}$ which requires (see e.g. \cite{peebles93})
that $\Omega_0h^2>0.052$.

Since the equation of  radial null geodesics is given by
\begin{equation}
d\eta=\pm d\chi,\end{equation} the comoving particle horizon radius is
(assuming a big-bang at $\eta=0$)
$\chi^{com}_{horizon}(\eta)=\eta$. Then, the fundamental
polyhedron has completely entered the horizon when
$\chi^{com}_{horizon}(\eta)=r_+$, that is when
\begin{equation}
\eta=\eta_{in}=r_+;\quad
z_{in}=\frac{2(\Omega_0^{-1}-1)}{\cosh{r_+}-1}-1.\label{zmoins}
\end{equation}
${\cal P}$ is still outside the horizon today if
\begin{equation}
z_{in}\leq0\Longleftrightarrow r_+\geq \arg\cosh{(2\Omega_0^{-1}-1)}.
\end{equation}
Once ${\cal P}$ has entered the horizon, it takes a time of order
$r_+$ for the TD to completely disappear (i.e. for the cosmic string
loops to decay, for the texture to unwind etc.) so that one  approximatively
has that
\begin{equation}
\eta_{disp}\simeq 2r_+;\quad z_{disp}\simeq\frac{2(\Omega_0^{-1}-1)}
{\cosh{2r_+}-1}-1.\label{tdisp}
\end{equation}
I now estimate the angle $\Theta$ under which the sphere of comoving
radius $r_+$ located at a redshift $z$ is seen. For that purpose, I
have to introduce the angular diameter distance $R_{_A}$ defined by the
relation (see e.g. \cite{ellis71b})
\begin{equation}
dS_0\equiv R_{_A}^2d\Omega_0,
\end{equation}
where $d\Omega_0$ is the solid angle substended by a bundle of null
geodesics diverging from the observer and $dS_0$ the cross-sectional
area of the bundle at the point that is observed. When the distortion is
not large (which the case in a Friedmann-Lema\^\i tre universe),
$R_{_A}$ is related to the observed angular diameter $\Theta$ of some
object whose comoving diameter $2r_+$ is perpendicular to the line of sight by
\cite{ellis71b}
\begin{equation}
\Theta(z)\simeq\frac{2r_+^{phys}}{R_{_{A}}(z)}=
(1+z)^{-1}\frac{2r_{+0}^{phys}}{R_{_{A}}(z)}
.\label{deftheta}
\end{equation}
In a pure matter universe, $R_{_A}(z)$ can be computed (see
e.g. \cite{peebles93,ellis71b}),
\begin{equation}
R_{_{A}}(z)=\frac{2\left[z+(1-2\Omega_0^{-1})(\sqrt{1+\Omega_0z}-1)\right]}
{H_0\Omega_0(1+z)^2}.\label{da}
\end{equation}
From (\ref{lphys}), (\ref{tdisp}), (\ref{deftheta}) and (\ref{da}), I
deduce that the angle under which the fundamental polyhedron is seen
at the time all the TD have disappeared is
\begin{equation}
\Theta_{disp}\simeq\frac{\Omega_0r_+}{\sqrt{1-\Omega_0}}\frac{1+z_{disp}}
{z_{disp}+(1-2\Omega_0^{-1})(\sqrt{1+\Omega_0z_{disp}}-1)}\label{theta}.
\end{equation}
This formula is the keystone of the following discussion.

\subsection{CMB anisotropies}

I now estimate the CMB anisotropies in models where the fluctuations
are generated by TD in a multi-connected universe at large angular
scales to see if they can reproduce their observed behaviour.

Neglecting the thickness of the last scattering surface, which is an
excellent approximation for the long wavelength modes, the temperature
anisotropies generated by the scalar modes. 
can be found by integrating the photon geodesics between
the last scattering surface and their reception (see e.g. \cite{panek86}),
\begin{equation}
\frac{\delta T}{T}(\vec n)=\left[\frac{1}{4}\delta_\gamma+
n^i\partial_iV +\Phi +{\cal H}V\right]_{_{LSS}}+\int_{_{LSS}}^0(\Psi'+\Phi')
d\eta.\label{SW}
\end{equation}
$\delta_\gamma$ and $V$ are respectively the photon density
contrast and the baryon velocity perturbation in comoving gauge,
$\Phi$ and $\Psi$ the two gravitational potentials in newtonian gauge
\cite{mukhanov92} and $\vec n$ the direction of observation. 
The first term will be refered to as the intrinsic Sachs-Wolfe term and
to the second term as the integrated Sachs-Wolfe term.

The multipole moments $C_\ell$ of the angular power spectrum of the
CMB anisotropies are related to the temperature two point correlation
function according to
\begin{equation}
\left<\frac{\delta T}{T}(\vec n_1)\frac{\delta T}{T}(\vec n_2)
\right>_{_{\vec n_1.\vec n_2=\cos{\theta}}}=
\sum_l\frac{(2\ell+1)}{4\pi}C_\ell P_\ell(\cos{\theta}),
\end{equation}
where the $P_\ell$ are the Legendre polynomials. The brackets denote an
average on the sky, i.e. on all pairs ($\vec n_1,\vec n_2$) such that
$\vec n_1.\vec n_2=\cos{\theta}$. Since the $\ell^{th}$ Legendre
polynomial has $2\ell$ zeros in $[-\pi,\pi]$ with approximatively equal
spacing, I can estimate that a parameter $\ell$ corresponds to an
angular scale of $\theta\simeq\pi/\ell$.

Observationally \cite{smoot96}, there is the so-called ``Sachs-Wolfe
plateau'' for small $\ell$, i.e. the $C_\ell$ behave as
\begin{equation}
C_\ell\sim A\ell^{-1}(\ell+1)^{-1},\label{COBE}
\end{equation}
A being a constant. I now focus on this plateau.

Given a model of generation of the cosmological perturbations, one can
compute the $C_\ell$ and compare with the observations. For that
purpose, all these perturbations should be decomposed on the
eigenfunctions $Q_{klm}$ of the Laplacian (see
e.g. \cite{langlois97}). I must emphasize here that in general these
eigenfunctions and eigenmodes are not known on an arbitrary topology
\cite{cornish98}. I will then make a ``{\it semi-topological}''
approximation consisting in taking into account the constraints from the
topology on the perturbations by the fact that the TD have
disappeared after $\eta_{disp}$ and treating the manifold as
simply-connected to compute the $C_\ell$.

Let us compare the different contributions in (\ref{SW}) for a TD
model in a ($\Omega_0<1$)-universe. On large scales, the integrated
Sachs-Wolfe term is the dominant contribution and has two origins. The
first one comes from the fact that the TD interact with the matter
between the last scattering surface and today, the second one
is specific to non-flat universes where the gravitational potentials
are not constant on superhorizon scales \cite{mukhanov92,kamionkowsky94}. 

In defect models, the Sachs-Wolfe plateau is recovered from the integrated
Sachs-Wolfe term \cite{durrer}. The time evolution of the
potentials [$\Phi'+\Psi'$ in equation (\ref{SW})] is driven by
the perturbations seeded by the defects between the last scattering
surface and us (see e.g. \cite{deruelle97} for a discussion of
the dominant terms).  In a simply-connected universe, the
process of decay and motion of the TD on the line of sight is
continuous up to now and enables to recover the behaviour
(\ref{COBE})\cite{durrer}. 

Now, in a multi-connected universe, there is no TD left after
$\eta_{disp}$, and thus no perturbations can be seeded on scales
larger than $\Theta_{disp}$. This induces an approximate cut-off in
the $C_\ell$ at
\begin{equation}
\ell_{min}(\Omega_0,r_+)\sim{\pi}/{\Theta_{disp}}.\label{lmin2}
\end{equation}
Using (\ref{theta}) and (\ref{tdisp}), the value of $\ell_{min}$ can
be estimated once the topology of the spatial sections is known,
i.e. $r_+$ (e.g. $r_+=0.7525$ for the Weeks manifold and $r_+=0.7485$
for the Thurston manifold) and the density parameter
$\Omega_0$. According to the value of ($\Omega_0,r_+$), it can be seen
from Fig. 1 wether TD can explain the Sachs-Wolfe plateau or not. The
existence of a cut-off $\ell_{min}$ is in obvious contradiction with
COBE observations (\ref{COBE}) when $\ell_{min}>1$.
\begin{figure}
\centering
\epsfig{figure=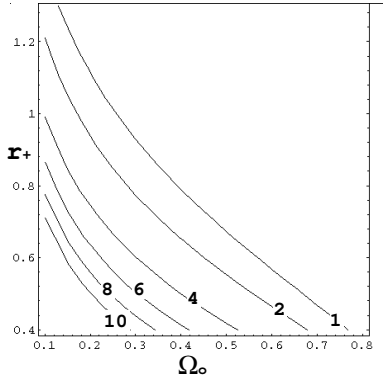, width=5cm, height=5cm}
\caption{Level lines $\ell_{min}=(1,2,4,6,8,10)$ computed from
(\ref{lmin2}) and (\ref{theta}) for the smallest CMB multipoles that can be
generated in a universe with density parameter $\Omega_0$ and
outradius $r_+$ in a theory where the perturbations are generated by
TD. Note that this can only be an approximate guide.\label{lmin}}
\end{figure}
\section{Conclusions}
In this article, I  have studied models where cosmological
perturbations are generated by TD in a multi-connected universe. Given
a cosmological model, i.e. ($\Omega_0,r_+$), it can be concluded 
(modulo the approximations I have used) from
Fig. 1 whether or not TD can account for the Sachs-Wolfe plateau
observed by COBE. In a wide class of multi-connected universes,
although TD may appear, they cannot explain COBE observations.

The conclusions rely on an important property of compact hyperbolic
3-manifolds (namely that their volume is a topological invariant,
contrary to e.g.  the 3-torus whose size is arbitrary) and on the
``{\it semi-topological}'' approximation which is a way to take into
the topological constraints. I also restricted my analysis to the
scalar modes, but the conclusion will be alike for vector and tensor
modes since they are not generated after the time $\eta_{disp}$.\\

I am grateful to N.~Deruelle for stimulating discussions and for helping
me to improve this text. I also wish to thank P.~Peter, D.~Langlois,
G.~Faye and N.~Cornish for their interesting remarks.

 \end{document}